\documentclass[aps,twocolumn,superscriptaddress]{revtex4}
\usepackage{amsmath,amssymb,epsfig,natbib,bm}

\def\be{\begin{equation}}
\def\ee{\end{equation}}
\def\bea{\begin{eqnarray}}
\def\eea{\end{eqnarray}}
\def\biposh#1#2#3{ A^{#1}_{#2|#3}}
\def\tildebiposhmat#1#2#3{ {\bf\tilde{A}}^{#1}_{#2|#3}}
\def\biposhmat#1#2#3{ {\bf A}^{#1}_{#2|#3}}
\begin{document}

\newcommand{\edth}{\,\eth\,}
\renewcommand{\beth}{\,\overline{\eth}\,}
\newcommand{\leftexp}[2]{{\vphantom{#2}}^{#1}{#2}}

\title{Statistical Isotropy of CMB Polarization Maps}

\author{Soumen Basak}
\affiliation{ Inter-University Centre for Astronomy and Astrophysics, Ganeshkhind, Post Bag 4, Pune-411007, India.}
\author{Amir Hajian}
\affiliation{ Inter-University Centre for Astronomy and Astrophysics, Ganeshkhind, Post Bag 4, Pune-411007, India.}
\affiliation{  Department of Physics, Jadwin Hall, Princeton University, PO Box 708, Princeton, NJ 08542. }
\affiliation{  Department of Astrophysical Sciences, Peyton Hall, Princeton University, Princeton, NJ 08544. }

\author{Tarun Souradeep}
\affiliation{ Inter-University Centre for Astronomy and Astrophysics, Ganeshkhind, Post Bag 4, Pune-411007, India.}
\begin{abstract}
We formulate statistical isotropy of CMB anisotropy maps in its most
general form. We also present a fast and orientation independent
statistical method to determine deviations from statistical isotropy
in CMB polarization maps. Importance of having statistical tests of
departures from SI for CMB polarization maps lies not only in
interesting theoretical motivations but also in testing cleaned CMB
polarization maps for observational artifacts such as residuals from
polarized foreground emission. We propose a generalization of the
Bipolar Power Spectrum (BiPS) to polarization maps. Application to the
observed CMB polarization maps will be soon possible after the release
of WMAP three year data. As a demonstration we show that for
E-polarization this test can detect breakdown of statistical isotropy
due to polarized synchrotron foreground.
\end{abstract}

\maketitle

\section{Introduction}

In a very near future we are going to have the first ``full'' sky CMB
polarization maps. The wealth of information in the CMB polarization
field will enable us to determine the cosmological parameters and test
and characterize the initial perturbations and inflationary mechanisms
with great precision. Cosmological polarized microwave radiation in a
simply connected universe is expected to be statistically
isotropic. This is a very important feature which allows us to fully
describe the field by its power spectrum that can have profound
theoretical implications for cosmology. Violation of statistical
isotropy (SI) in CMB polarization maps is going to be very important
soon. It can now be tested with CMB polarization maps over large sky
fraction. Importance of having statistical tests of departures from SI
for CMB polarization maps lies not only in interesting theoretical
motivations but also in testing the cleaned CMB polarization maps for
residuals from polarized foreground emission. Unlike the foregrounds
in temperature anisotropies, polarized foreground emissions on large
scales. In these scales we expect to see the primordial $B$-mode due
to inflationary gravitational waves at all frequencies. A robust
discriminator between the primordial polarized radiation and polarized
foreground emissions is the test of SI. In this paper we study
statistical isotropy in its most general form based on the
Bipolar Power Spectrum (BiPS) that was proposed as a measure of SI
violation in CMB temperature \cite{us_apjl, us_pramana, us_jgrg}. The
BiPS has been applied to check for the SI of CMB temperature maps
based on the WMAP first year data \cite{us_pramana, us_jgrg}. We
present a simple formalism that works for all three scalar fields that
describe CMB temperature and polarization, $T$, $E$ and $B$. Then we
use BiPS as a diagnostic tool to check for departures from SI in $E$
and $B$ polarization modes as well as the cross terms such as $TE$. We
present an example of applying the method on simulated CMB
polarization maps that include polarized foreground from the
synchrotron emission in our galaxy.

The rest of this paper is organized as follows: section II is a very
brief introduction to polarization and temperature anisotropy of CMB
and shows how CMB anisotropy can be fully described by three scalar
fields, $T$, $E$ and $B$.  Section III is dedicated to formulation of
statistical isotropy in general. Section IV defines an unbiased
estimator for bipolar power spectrum (BiPS) which is shown to be a
strong tool for testing departures from statistical isotropy in a
given map. And finally section V describes an example of how this
method works for a E-polarization where statistical isotropy is
violated due to large galactic foreground from synchrotron
emission. We provide some useful mathematical relations in
the appendix.

\section{CMB Anisotropy and Polarization Maps}
CMB anisotropy is completely described by its temperature anisotropy, and
polarization. Temperature anisotropy is a scalar random field, $\Delta
T(\hat{n})=T(\hat{n})-T_0$, on a 2-dimensional surface of a sphere
(the sky), where $\hat{n}=(\theta,\phi)$ is a unit vector on the
sphere and $T_0=\int{\frac{d\Omega_{\hat{n}}}{4\pi}T(\hat{n})}$
represents the mean temperature of the CMB. It is convenient to expand
the temperature anisotropy field into spherical harmonics, the
orthonormal basis on the sphere, as \be \label{yelemexpand} \Delta
T(\hat{n}) \, = \, \sum_{l,m} a_{lm}Y_{lm}(\hat{n}) \,\,, \ee where
the complex quantities, $a_{lm}$ are given by \be \label{alm} a_{lm} =
\int{\mathrm{d} \Omega_{\hat{n}}Y_{lm}^{*}(\hat{n}) \Delta
T(\hat{n})}.  \ee

CMB polarization field is described by the Stokes parameters,
$Q(\hat{n})$ and $U(\hat{n})$, which depend on the choice a local
Cartesian patch the coordinate on the sky. One can combine
these Stokes parameters into two complex quantities, $Q(\hat{n}) -
i~U(\hat{n})$ and $Q(\hat{n})+ i~U(\hat{n})$ which transform like
spin-2 fields under rotations of the coordinates by an angle $\psi$,
\be (Q(\hat{n})\pm i~U(\hat{n}))'=e^{\mp 2i\psi} (Q(\hat{n})\pm
i~U(\hat{n})).  \ee One may thus expand each of them in terms of
spin-weighted spherical harmonics, ${}_{\pm 2}Y_{l m}$,
\begin{eqnarray}
Q(\hat{n}) - i~U(\hat{n}) &=& \sum_{l m} a_{2,l m}\,{}_{2}Y_{l m}(\hat{n})\\ \nonumber
Q(\hat{n}) + i~U(\hat{n}) &=& \sum_{l m} a_{-2,l m}\,{}_{-2}Y_{l m}(\hat{n}) 
\end{eqnarray}

Applying spin-lowering (spin-raising) operators $\beth$ ($\edth$)
twice on ${}_{\pm 2}P(\hat{n})=Q(\hat{n}) \mp i~U(\hat{n})$ one can
construct two spin-zero fields,
\begin{eqnarray}
\beth^2_{\hat{n}} \,\,\,{}_{2}P(\hat{n})
&=& \sum_{lm} \left[\frac{(l + 2)!}{(l -2)!}\right]^{1/2}\, a_{2,l m}\,
 Y_{l m} (\hat{n})
 \nonumber \\
 \edth^2_{\hat{n}} \,\,\,{}_{-2}P(\hat{n})
&=& \sum_{lm} \left[\frac{(l + 2)!}{(l -2)!}\right]^{1/2}\, a_{-2,l m}\, Y_{l m}(\hat{n})
\end{eqnarray}
For fullsky maps, the above spin-2 fields can be linearly combined to
construct two scalar fields \cite{Zaldarriaga:1996xe,
Kamionkowski:1996ks}
\begin{eqnarray}
E(\hat{n}) &=& \frac{1}{2}\left[\beth_{\hat{n}}^2 \, {}_{2}P(\hat{n}) + \edth_{\hat{n}}^2  \, {}_{-2}P(\hat{n}) \right]\\ \nonumber
B(\hat{n}) &=& \frac{1}{2i}\left[\beth_{\hat{n}}^2  \, {}_{2}P(\hat{n}) - \edth_{\hat{n}}^2  \, {}_{-2}P(\hat{n}) \right].
\end{eqnarray}
Now, expanding these in terms of spherical harmonics,
\begin{equation}
E(\hat{n}) = \sum_{l m} a^{E}_{l m} \,Y_{l m}(\hat{n}); \,\,\,B(\hat{n})
= \sum_{l m} a^{B}_{l m}\,Y_{l m}(\hat{n})
\label{EB}
\end{equation}
we get,
\begin{equation}
a^{E}_{l m} = \frac{1}{2} \left(a_{2,l m} + a_{-2,l
m}\right); \,\,\,a^{B}_{l m} = \frac{1}{2i} \left(a_{2,l m} - a_{-2,l
m}\right)
\end{equation}
Therefore one can characterize CMB anisotropy in the sky maps by three
scalar random fields: $T(\hat{n})$, $E(\hat{n})$, and $B(\hat{n})$
with no loss of information. For cut-sky, $E(\hat{n})$ and $B(\hat{n})$
mode decomposition is not unique \cite{Lewis:2002,Brown:2004}. But
since mixing is linear there always exist two linearly independent
modes. It is possible to formulate the SI of these linear independent
modes.
Statistical properties of each of these fields can be characterized by
$N$-point correlation functions, $\langle X(\hat{n}_1)
X(\hat{n}_2)\cdots X(\hat{n}_n)\rangle$.
 Here the bracket denotes the ensemble average, {\it{i.e.}} an average over all
possible configurations of the field, and $X(\hat{n})$ can be any of
the $T(\hat{n})$, $E(\hat{n})$, or $B(\hat{n})$ fields. CMB anisotropy
is believed to be Gaussian \cite{Bartolo:2004, Komatsu:2003}. Hence
the connected part of $N$-point functions disappears for $N >
2$. Non-zero (even-$N$)-point correlation functions can be expressed
in terms of the $2$-point correlation function. As a result, a
Gaussian distribution is completely described by two-point correlation
functions of $X(\hat{n})$, \be
C^{XX'}(\hat{n},\hat{n'})\,=\, \langle X(\hat{n}) X'(\hat{n}')
\rangle.  \ee 
Equivalently, as it is seen from linear relations in
eqns.~(\ref{alm}) and (\ref{EB}), for a Gaussian CMB anisotropy,
$a^X_{lm}$ are Gaussian random variables too. Therefore, the {\it
covariance matrix}, $\langle a^X_{lm}a^{X'*}_{l^\prime
m^\prime}\rangle$, fully describes the whole field. Throughout this
paper we assume Gaussianity to be valid.

\section{Statistical isotropy}
Two point correlations of CMB anisotropy, $C^{XX'}(\hat{n}_1,\,
\hat{n}_2)$, are two point functions on $S^2 \times S^2$, and hence
can be expanded as \be \label{bipolar} C^{XX'}(\hat{n}_1,\,
\hat{n}_2)\, =\, \sum_{l_1,l_2,\ell,M} \biposh{XX'}{\ell M}{l_1 l_2}
Y^{l_1l_2}_{\ell M}(\hat{n}_1,\, \hat{n}_2).
\ee Here $\biposh{XX'}{\ell M}{l_1 l_2}$ are coefficients of the
expansion (hereafter BipoSH coefficients) and $Y^{l_1l_2}_{\ell
M}(\hat{n}_1,\, \hat{n}_2)$ are bipolar spherical harmonics defined by
eqn.~(\ref{bipolars}). Bipolar spherical harmonics form an orthonormal
basis on $S^2 \times S^2$ and transform in the same manner as the
spherical harmonic function with $\ell,\, M$ with respect to rotations
\cite{Var}. We can inverse-transform $C^{XX'}(\hat{n}_1,\,
\hat{n}_2)$ in eqn.~(\ref{bipolar}) to get the coefficients of
expansion, $\biposh{XX'}{\ell M}{l_1 l_2}$, by multiplying both sides of
eqn.(\ref{bipolar}) by $Y^{*l'_1l'_2}_{\ell'M'}(\hat{n}_1,\hat{n}_2)$
and integrating over all angles. Then the orthonormality of bipolar
harmonics, eqn. (\ref{A2}), implies that \be \label{alml1l2}
\biposh{XX'}{\ell M}{l_1 l_2} \,=\,\int d\Omega_{\hat{n}_1}\int
d\Omega_{\hat{n}_2} \, C^{XX'}(\hat{n}_1,\, \hat{n}_2)\,
Y^{*l_1l_2}_{\ell M}(\hat{n}_1,\hat{n}_2).  \ee The above expression
and the fact that $C^{XX}(\hat{n}_1,\, \hat{n}_2)$ is symmetric under
the exchange of $\hat{n}_1$ and $\hat{n}_2$ lead to the following
symmetries of $\biposh{XX}{\ell M}{l_1 l_2}$ \bea \label{sym}
\biposh{XX}{\ell M}{l_2 l_1} \,&=&\,(-1)^{(l_1+l_2-L)}\biposh{XX}{\ell M}{l_1 l_2}, \\ \nonumber \biposh{XX}{\ell M}{ll} \, &=& \, \biposh{XX}{\ell M}{ll}
\,\,\delta_{\ell,2k-1}, \,\,\,\,\,\,\,\,\,\,\,\,\,\,\,\,\,\,\,\,\,
k=1,\,2,\,3,\,\cdots.  \eea

It has been shown \cite{us_bigpaper} that Bipolar Spherical Harmonic
(BipoSH) coefficients, $\biposh{XX'}{\ell M}{l_1 l_2}$, are in fact linear
combinations of off-diagonal elements of the covariance matrix, \be
\label{ALMvsalm} \biposh{XX'}{\ell M}{l_1 l_2} \,=\, \sum_{m_1m_2} 
\langle a^X_{l_1m_1}a^{*X'}_{l_2 m_2}\rangle (-1)^{m_2} {\mathcal
  C}^{\ell M}_{l_1m_1l_2 -m_2}.  \ee where ${\mathcal
  C}_{l_1m_1l_2m_2}^{\ell M}$ are Clebsch-Gordan coefficients. This
  clearly shows that $\biposh{XX'}{\ell M}{l_1 l_2}$ completely
  represent the information of the covariance matrix. When statistical
  isotropy holds, it is guaranteed that the covariance matrix is
  diagonal, \be \langle a^X_{lm}a^{*X'}_{l' m'}\rangle = C^{XX'}_{l}
  \delta_{ll^\prime} \delta_{mm'} \ee and hence the angular power
  spectra carry all information of the field. Substituting this into
  eqn.~(\ref{ALMvsalm}) gives \be \label{SIALM1} \biposh{XX'}{\ell
  M}{ll'} \,=\,(-1)^l C^{XX'}_{l} (2l+1)^{1/2} \, \delta_{ll^\prime}\,
  \delta_{\ell 0}\, \delta_{M0}.  \ee The above expression tells us
  that when statistical isotropy holds, all BipoSH coefficients,
  $\biposh{XX'}{\ell M}{ll'}$, are zero except those with $\ell=0,
  M=0$ which are equal to the angular power spectra up to a $(-1)^l
  (2l+1)^{1/2}$ factor. BipoSH expansion is the most general way of
  studying two point correlation functions of CMB anisotropy. The well
  known angular power spectrum, $C_l$ is in fact a subset of the
  corresponding BipoSH coefficients, \be \label{SIALM}
  C^{XX'}_{l}\,=\, \frac{(-1)^{l}}{\sqrt{2l+1}} \biposh{XX'}{0 0}{ll}.
  \ee Therefore to test a CMB map for statistical isotropy, it is
  enough to compute the BipoSH coefficients for the maps and check for
  nonzero BipoSH coefficients. Every statistically significant
  deviation of BipoSH coefficients from zero would mean violation of
  statistical isotropy. In the next section we discuss this in more
  details.

\section{Unbiased Estimator}
In statistics, an estimator is a function of the known data that is
used to estimate an observable quantity. An estimate is the result of
the actual application of the function to a particular set of
data. Different estimators may be defined for a given observable. The
above theory can be used to construct an estimator for measuring
BipoSH coefficients from a given CMB map as,
\be
\label{estimator}
\biposh{XX'}{\ell M}{ll'} =\sum_{m m^\prime} \sqrt{W_l W_{l'}}\,
a^X_{lm}a^{X'}_{l^\prime m^\prime} \, \, {\mathcal{ C}}^{\ell
M}_{lml^\prime m^\prime}\,\quad , \ee where $W_l$ is the Legendre
transform of the window an isotropic smoothing function that can be
applied to the data. The ensemble average of this estimator is given
by, \be \left<\biposh{XX'}{\ell M}{ll'}\right> = \sum_{m m^\prime}
\sqrt{W_l W_{l'}}\, \left<a^X_{lm}a^{X'}_{l^\prime m^\prime}\right> \,
\, {\mathcal{ C}}^{\ell M}_{lml^\prime m^\prime}\,\quad , \ee
which is its true value. Akin to the well known quadratic estimator
$\hat{C}_l=\frac{1}{2l+1}\sum_{m}{|a_{lm}|^2}$ for $C_{l}$, the above
estimator is an unbiased estimator of BipoSH coefficient.
However it is impossible to measure all $\biposh{XX'}{\ell
M}{ll'}$ individually because of cosmic variance. Combining BipoSH
coefficients helps to reduce the cosmic variance.
Among the several possible combinations of BipoSH coefficients, the
Bipolar Power Spectrum (BiPS) has proved to be a useful tool with
interesting features. BiPS of CMB anisotropy is defined as a quadratic
contraction of the BipoSH coefficients \be
\label{kappal} \kappa_\ell^{XX'} \,=\, \sum_{l,l',M}
\left|\biposh{XX'}{\ell M}{ll'}\right|^2 \geq 0.  \ee Some interesting
properties of BiPS are as follows: it is orientation independent, {\it
i.e.} invariant under rotations of the sky. For models in which
statistical isotropy is valid, BipoSH coefficients are given by
eqn~(\ref{SIALM}), and therefore lead to a null BiPS, {\it i.e.}
$\kappa_\ell\,=\,0$ for every $\ell>0$, \be
\kappa_\ell^{XX'}\,=\,\kappa_0 \delta_{\ell 0}.  \ee Non-zero
components of BiPS imply break down of statistical isotropy, and this
introduces BiPS as a measure of statistical isotropy, \be {\mathrm
{\Huge STATISTICAL\,\,\,\, ISOTROPY}} \,\,\,\,\,\,\, \Longrightarrow
\,\,\,\,\,\,\, \kappa_\ell\,=\,0 \,\,\,\,\,\,\, \forall \ell \ne 0.
\ee It is important to note that although BiPS is quartic in
$a_{lm}$, it is designed to detect SI violation and not
non-Gaussianity \cite{us_bigpaper, us_apjl, us_pramana, us_jgrg,
us_apj}. An un-biased estimator of BiPS is given by \be
\label{estimatork}
\tilde\kappa_\ell^{XX'} = \sum_{ll^\prime M}
\left|\biposh{XX'}{\ell M}{ll'}\right|^2 - {\mathfrak
B}_\ell^{XX'}\, , \ee where ${\mathfrak B}_\ell^{XX'}$ is the bias
related to the SI part of the map and given by the angular
power spectrum, $C_l$, 
\bea
\label{klisobias}
{\mathfrak B}_\ell^{XX'} &\equiv&
\langle\tilde\kappa_\ell^B\rangle_{_{\rm SI}} = (2\ell+1)\,\sum_{l_1}
\sum_{l_2=|\ell-l_1|}^{\ell+l_1} W_{l_1} W_{l_2} \times  \nonumber \\&
& \,\,\,\,\,\left[ C^{XX}_{l_1} C^{X'X'}_{l_2} +
(-1)^{\ell}\, \delta_{l_1 l_2} \left(C^{XX'}_{l_1}\right)^{2}
\right]\,.  \eea The above expression for ${\mathfrak B}_\ell$ is
obtained by assuming Gaussian statistics of the temperature
fluctuations \cite{us_apjl, us_bigpaper}.  {\em Note, the estimator
$\tilde \kappa_\ell$ is unbiased, only for SI correlation. In that
case, ensemble average of $\tilde \kappa_\ell$ is same as its true
value which is zero for $\ell \ne 0$, i.e., $\langle \tilde
\kappa_\ell \rangle=0$.}


\section{Example: Polarized Synchrotron Contamination}
\begin{figure}[h]
\includegraphics[scale=0.40, angle=0]{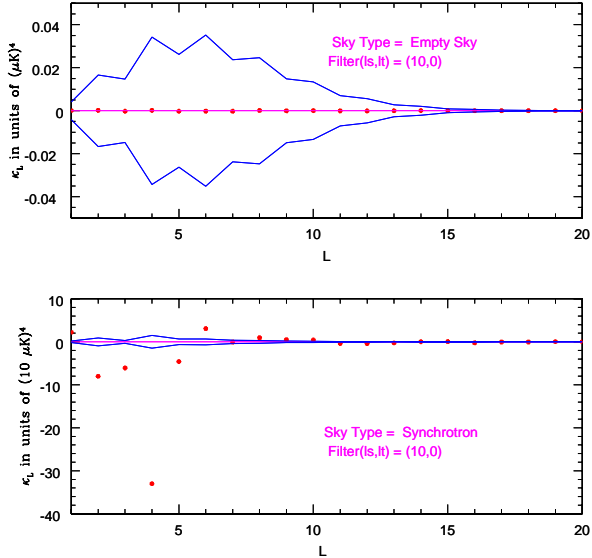}

  \caption{CMB polarization maps with no foregrounds are statistically
  isotropic and have null bipolar power spectrum (top). Adding
  polarized synchrotron emission violates statistical isotropy and
  results in a detectable non-zero bipolar power spectrum
  (bottom). Red dots show the BiPS after bias subtraction and Blue
  lines show the  $1-\sigma$ of the cosmic variance.  }
  \label{results1}
\end{figure}
\begin{figure}[h]
 \includegraphics[scale=0.40, angle=0]{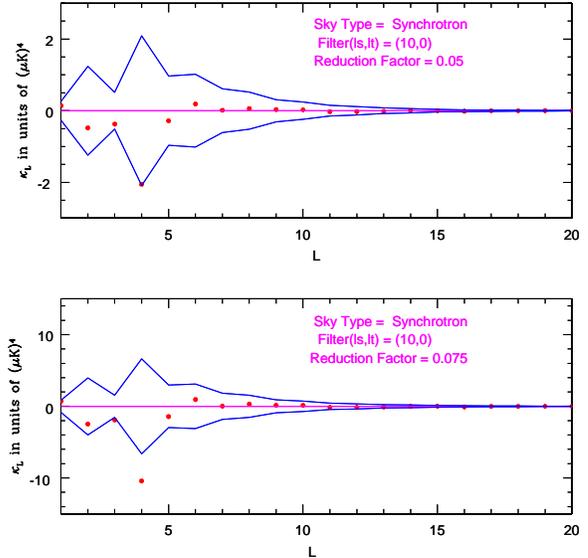}
  \caption{Adding $5\%$ of the polarized synchrotron emission just
  violates statistical isotropy (top). Adding $7.5\%$ of the polarized
  synchrotron emission clearly violates statistical isotropy and
  results in a detectable non-zero bipolar power spectrum
  (bottom). Red dots show the BiPS after bias subtraction and Blue
  lines show the $1-\sigma$ of the cosmic variance.  }
  \label{results2}
\end{figure}
As an example of how one can detect deviations from statistical
isotropy in CMB polarization maps, we make statistically anisotropic
polarization maps and estimate the BiPS from them. This can be done in
many different ways but here we choose a simple method which results
in severe violation of SI and therefore is good for a demonstration of
the method. We add  polarized synchrotron emission template to the
background CMB polarization map. Polarized synchrotron template 
$(30~GHz)$ is made using Planck Simulator \cite{plancksimulator} which uses
the model by \cite{Giardino:2002}, {\it i.e.} the polarization degree
is a function of the intensity spectral index while polarization
angles are derived from a gaussian distribution. Here we restrict our
attention to $E$ mode polarization only. It is obvious that everything
can be done in the same way for B mode as well. The estimator will
then be \be
\label{estimatorE} \biposh{EE}{\ell M}{ll^\prime} = \sum_{m m^\prime}
\sqrt{W_l W_{l'}} a^E_{lm}a^E_{l^\prime m^\prime} \, \, {\mathcal{
C}}^{\ell M}_{lml^\prime m^\prime}\,\quad , \ee where $a^E_{lm}$ are
the spherical harmonic transform of the background CMB polarization
map plus the polarized synchrotron radiation, 
\be a^{E}_{l m} =
a^{E_{cmb}}_{lm} + a^{E_{sync}}_{lm}.  \ee 
and $W_{l}$ is a isotropic filter that allows us to target angular
scales of interest by filtering out power on other scales.

We simulate 1000 statistically isotropic CMB polarization maps, add
the synchrotron template to each of them and compute the BiPS for them
using the estimators of eqns.~(\ref{estimatorE}) and
(\ref{estimatork}). Filters that we use here can be divided into two
categories: low pass Gaussian filters \be W_{l}^{G} = N^{G}
\exp\left\{-\left(\frac{2l + 1}{2l_{s} + 1} \right)^{2}\right\}
\label{lowpass}\ee that cuts power on scales, $l \le l_{s}$ and band
pass filters of the form \be W_{l}^{S} = 2 N^{S} \left[ 1 -
J_{0}\left(\frac{2l + 1} {2l_{s} + 1}
\right)\right]\exp\left\{-\left(\frac{2l + 1}{2l_{s} + 1}
\right)^{2}\right\}\,, \label{bandpass}\ee that retains power on
scales $l_{s} \le l \le l_{t}$, where $J_0$ is the spherical bessel
function and $N^{G}$ and $N^{S}$ are normalization constants chosen
such that, $\sum_{l} \frac{(2l + 1)W_{l}}{2l(l + 1)} = 1 $, i.e., unit
rms for unit flat band angular power spectrum, $l(l + 1)\,C^{XX'}_{l} =
2\pi$.

Results of this computation are shown in Fig. \ref{results1} and
Fig. \ref{results2}. We see that CMB polarization maps with no
foregrounds are statistically isotropic and have null bipolar power
spectrum. Adding polarized synchrotron emission violates statistical
isotropy at large angular scales and results in a detectable non-zero
BiPS. Retaining only $5\%$ of the polarized
synchrotron emission just violates statistical isotropy at the
threshold of $1-\sigma$. At $7.5\%$ of the polarized synchrotron
emission clearly shows the violation of statistical isotropy and results
in a sharply detectable non-zero bipolar power spectrum at $\kappa_{4}$.

We should emphasize that this is simply an example to demonstrate how
violation of statistical isotropy can be quantified in CMB
polarization maps. In reality, we usually expect to deal with cleaned
polarized maps which would contain some residuals that have different
angular structure. The signal would be much weaker and also have
different BiPS characteristics. Hunting tiny residuals from
foregrounds in maps of temperature anisotropy using statistical
isotropy has been studied \cite{us_foregrounds} and similar strategy
can be applied to polarization maps when they are available. In
addition, other observational artifacts such as anisotropic noise or
incomplete (masked) sky can also cause violation of statistical
isotropy in a polarization map. In the latter case, the incomplete sky
coverage immediately induces a contamination of E-mode of polarization
by its B-mode and vice-versa. Then the modified temperature and
polarization fields is related to their actual values of full sky
coverage by a window matrix \cite{Lewis:2002,Brown:2004} whose
elements are basically window functions for temperature and
polarization in harmonic space. It can be shown that the estimated
BipoSH coefficients are in fact linear combinations of that for
fullsky CMB maps.  
\be \tildebiposhmat{}{\ell M}{l l'} = \sum_{\ell'
M' l_1 l_2} {\bf N}^{\ell M l l'} _{\ell' M' l_1 l_2}\,\,
\biposhmat{}{\ell M}{l_1 l_2}\,. \label{biposhmat}\ee 
Here bold-faced $ \tildebiposhmat{}{\ell M}{l l'}$ and
$\biposhmat{}{\ell M}{l_1 l_2}$ are the column matrices corresponding
to estimated and true BipoSH coefficients respectively, for the auto
and cross-correlations $(TT, TE, TB, ET, EE, EB, BT, BE, BB)$ of
temperature anisotropy and polarization. The elements of the matrix $
{\bf N}^{\ell M l l'}_{\ell' M' l_1 l_2}$ depend on Clebsch- Gordan
coefficients and window functions in harmonic space. Hence, the true BipoSH
coefficients can be estimated from the pseudo-BipoSH coefficients by
inverting the above equation. We defer this to a future publication, a
SI analysis of CMB polarization when this effect is
important. However, we have verified using simulations that the BiPS
of polarization maps is insensitive to breakdown of SI due to galactic
cut when it is filtered at low-l using, $W_{l}^{G}(l_s=10,l_t=0)$ and
$W_{l}^{S}(l_s=20,l_t=10)$ in eqns.(\ref{lowpass}) \&
(\ref{bandpass}). (This is consistent with the result for cut-sky CMB
temperature maps discussed in the paper \cite{us_bigpaper}.) As a
result, the BiPS signature of the polarized galctic foregrounds
presented here would not change if the maps are masked by a galactic
cut.
CMB polarization maps filtered with windows peaked at higher
multipoles (eg., $W_{l}^{S}(l_s=90,l_t=80)$) do reflect the SI
violation arising from a galactic cut.  The complications of
quantifying statistical isotropy in cut-sky polarization CMB maps is
formally encoded by eqn.~(\ref{biposhmat}) but its implementation is a
challenging task which is currently under progress. (The effects can
also be estimated through extensive simulations.)

\section{Summary}

We present a novel approach to quantify the violation of statistical
isotropy in CMB polarization maps for the first time. We present a
fast and orientation independent method which allows for a general
test of isotropy using Bipolar Power Spectrum. This method has been
previously applied to the temperature anisotropy maps and many various
aspects of that are well studied in details. In this paper we extend
BiPS to the CMB polarization maps and by present a working example to
demonstrate its potential.

\acknowledgments We acknowledge fruitful discussions with Dmitri
Pogosyan, Lyman Page, David Spergel and Joe Silk. Our computations
were performed on Hercules, the high performance facility of IUCAA.
AH acknowledges support from NASA grant LTSA03-0000-0090. Some of the
analysis in this paper used the HEALPix package. We acknowledge
the use of the Legacy Archive for Microwave Background Data Analysis
(LAMBDA). Support for LAMBDA is provided by the NASA Office of Space
Science.

\appendix
\section{Useful mathematical relations}
Bipolar spherical harmonics form an orthonormal basis of $S^2 \times
S^2$ and are defined as \be
\label{bipolars}
Y^{l_1l_2}_{\ell M}(\hat{n}_1,\, \hat{n}_2)\,=\, \sum_{m_1m_2}
{\mathcal C}_{l_1m_1l_2m_2}^{\ell M} Y_{l_1 m_1}(\hat{n}_1)Y_{l_2
m_2}(\hat{n}_2), \ee in which ${\mathcal C}_{l_1m_1l_2m_2}^{\ell M}$
are Clebsch-Gordan coefficients. Clebsch-Gordan coefficients are
non-zero only if triangularity relation holds, $\{l_1l_2\ell\}$, and
$M=m_1+m_2$. Where the $3j$-symbol $\{abc\}$ is defined by
\begin{displaymath}
\{abc\} = \left\{
\begin{array}{ll}
1 &\textrm{if $a+b+c$ is integer and $|a-b|\leq c \leq (a+b)$,}  \\
0 &\textrm{otherwise,} 
\end{array} \right.
\end{displaymath}
Orthonormality of bipolar spherical harmonics \be \int
d\Omega_{\hat{n}_1}d\Omega_{\hat{n}_2} \, Y^{l_1l_2}_{\ell
M}(\hat{n}_1,\, \hat{n}_2) Y^{*l'_1l'_2}_{\ell' M'}(\hat{n}_1,\,
\hat{n}_2) = \delta_{l_1l'_1} \delta_{l_2l'_2}\delta_{\ell
\ell'}\delta_{MM'} \label{A2} \ee

\end{document}